\documentclass[10pt]{blois07}
\usepackage{graphicx}
\usepackage{cite,./mcite}

\begin{document}
\title{Blois07/EDS07 \\ Proceedings }
\author{Andrew Hamilton$^1$}
\institute{$^1$D\'epartment de physique nucl\'eaire et corpusculaire (DPNC), Universit\'e de Gen\`eve\\ e-mail: Andrew.Hamilton@cern.ch}
\maketitle
\begin{abstract}
The LHC will soon provide proton-proton collisions at the unprecedented center of mass energy, \mbox{$\sqrt{s}=$14~TeV}.  This not only allows us to probe new regions of high-$p_T$ physics, but also low-$x$ and forward physics.  A selection of potential measurements are described to outline the prospects for low-$x$ and forward physics in the ATLAS, CMS, TOTEM, and LHCf experiments.
\end{abstract}

\section{Prospects of Forward Energy Flow and Low-x Physics at the LHC}
\subsection{Introduction}
\label{sec:intro}

In a $pp$ collisions, forward collisions occur when $x_1 << x_2$, where $x_i$ is the fraction of the proton's total momentum carried by parton $i$ (the Bjorken-$x$).  Since one of the partons in a forward collision must have a low-$x$, there is an inherent connection between forward energy flow and low-x physics.  After a discussion on the current status of the forward detectors in the ATLAS, CMS, TOTEM, and LHCf experiments, some examples of the forward energy flow and low-$x$ physics potential of these experiments are given.

\begin{figure}[htbp]
   \centering
   \includegraphics[width=3in]{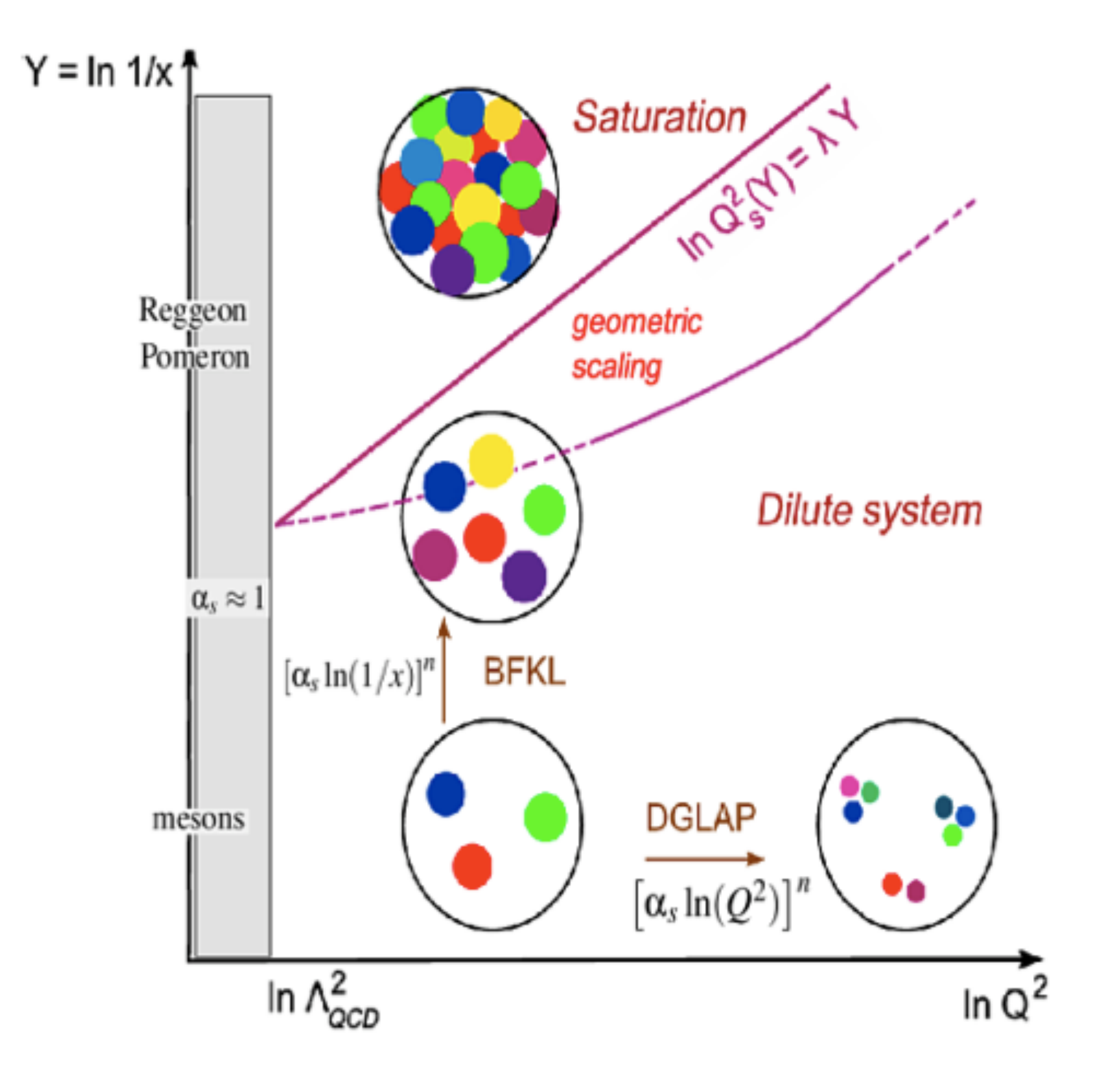}
   \includegraphics[width=2.5in]{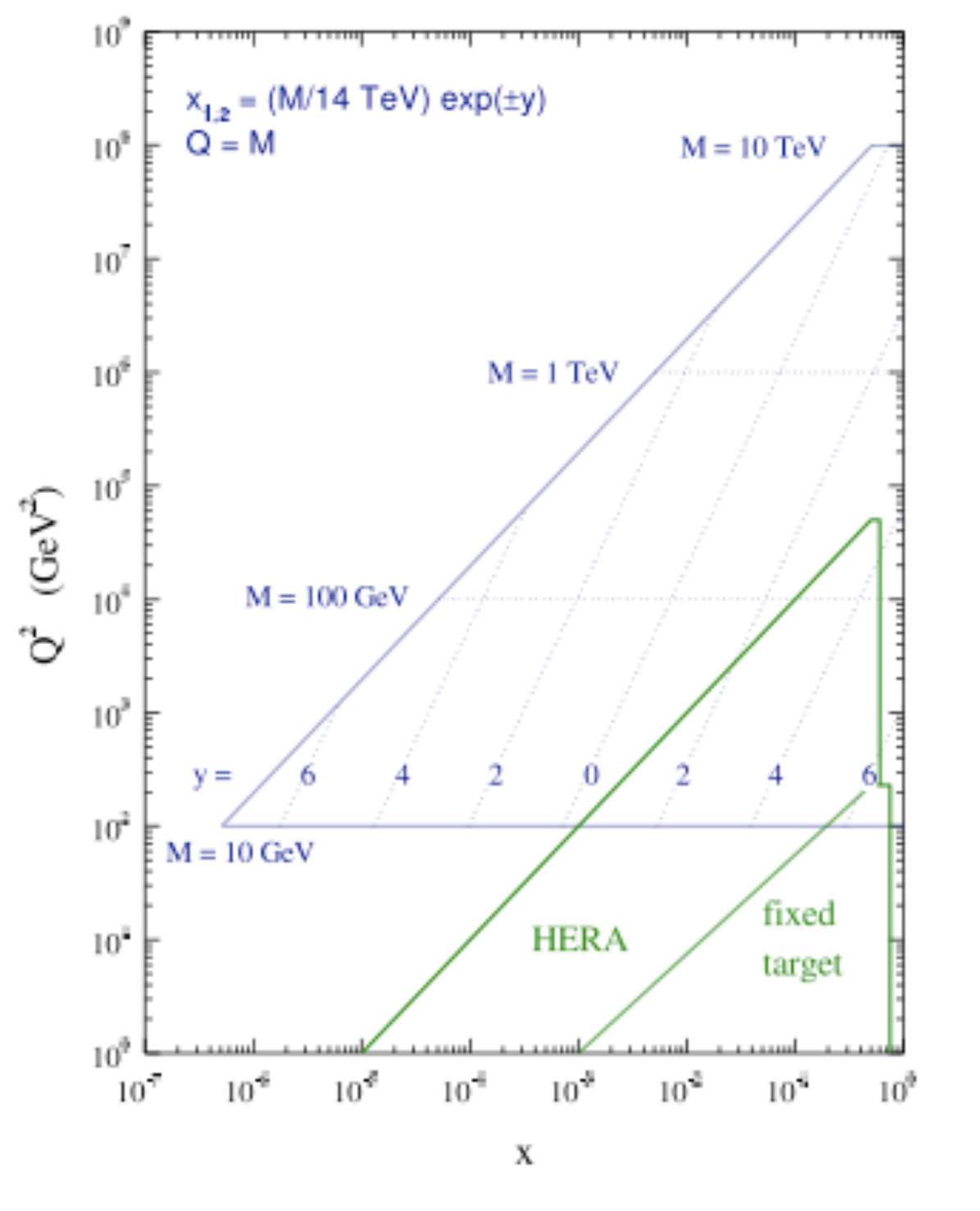} 
   \caption{Schematic of saturation region (left), LHC coverage in $x$ vs. $Q^2$ compared to HERA and fixed target experiments (right)}
   \label{fig:sat}
\end{figure}

The gluon density of the parton distribution functions (PDF) of the proton were discovered to grow as $xG(x,Q^2) \propto x^{-\lambda(Q^2)}$, where $\lambda \simeq 0.1 - 0.3$ rises logarithmically with $Q^2$, in deep-inelastic scattering (DIS) $ep$ collisions at HERA \cite{h1:pdf}.  The 'saturation' region, where the non-linear effects of gluon-gluon fusion due to high gluon density become important, is expected to be accessible at the LHC.  Figure~\ref{fig:sat} shows the coverage of the LHC in $x$ and $Q^2$~\cite{stirling:xvsq}, as well as a schematic representation of BFLK~\cite{lipatov:bfkl, kuraev:bfkl, balitsky:bfkl} evolution into the saturation region. 

Forward physics is also very interesting for the measurement of the high energy ($10^{16}-10^{20}$~eV) cosmic ray energy and composition spectrum.  Above $E_{lab} \sim 10^{14}$~eV only indirect measurements of extensive air showers are possible.  These indirect measurements depend on simulations of the air shower high in the atmosphere.  Since there is no accelerator data above Tevatron ($\sim10^{15}$~eV), the simulations have to extrapolate over several orders of magnitude to reach the highest energy cosmic rays at $\sim 10^{20}$~eV.  The dominant contribution to the uncertainty of these models is the soft QCD forward energy flow, which can be studied at the LHC.

\subsection{LHC Forward Detectors}
\label{sec:detectors}

The Large Hadron Collider (LHC) will provide $pp$ collisions with $\sqrt{s}=14$~TeV to 4 interaction points (IP) instrumented with detectors.  This is a summary of the forward detectors instrumented at IP1 and IP5.  The ATLAS and LHCf experiments are located at IP1, the CMS and TOTEM experiments are located at IP5.  The ALICE and LHCb experiments will not be discussed here.

The ATLAS detector~\cite{atlas:tdr} is the general purpose detector at IP1.  The ATLAS forward detectors  include LUCID, ALFA, and ATLAS-ZDC.  The LUCID and ALFA detector's~\cite{atlas:fwdloi} primary objective is luminosity measurement, but they will also be useful for forward physics studies.  The LUCID (\emph{LUminosity measurement using Cerenkov Integrating Detector}) detector is located at $\pm17$~m from the ATLAS IP ($5.4<|\eta|<6.1$) and consists of aluminium tubes filled with $C_4 F_{10}$ gas surrounding the beam-pipe and pointing towards the ATLAS IP.  The Cerenkov light emitted by a traversing particle from the ATLAS IP is reflected down the tube and read out by photo-multipliers.  Particles not coming from the ATLAS IP will not traverse an entire tube and thus leave a much smaller signal, thereby reducing backgrounds.  The LUCID detector is also expected to be of use in diffractive physics studies (using rapidity gap signatures) and forward multiplicity studies.  

The ALFA (\emph{Absolute Luminosity For ATLAS}) detectors are scintillating fibre trackers located inside roman pots at $\pm240$~m from the ATLAS IP.  The main purpose of ALFA is to measure elastic proton scattering at low angles to determine absolute luminosity in ATLAS.  To achieve an optimal precision in the luminosity measurement, the LHC will have dedicated runs with the so-called high-$\beta^{*}$ optics.  This will provide the absolute calibration of the luminosity for the LUCID detector.  Potential for other physics using ALFA, such as, measuring the total $pp$ cross-section, measuring elastic scattering parameters, and using the tagged protons for diffractive studies are also being explored.   

The ATLAS-ZDC (\emph{Zero Degree Calorimeter})~\cite{atlas:zdcloi} is a tungsten/quartz calorimeter located between the two LHC beam-pipes at $\pm140$~m from the ATLAS IP.  By measuring neutral particles at a $0^{\circ}$ polar angle it will have a central role in the ATLAS heavy-ion physics 
program, where it will be used to measure the centrality of the collisions, the luminosity, as well 
as be used in some physics triggers.   In the $pp$ physics program, it will be used to study forward particle production and energy flow.

The other experiment at IP1 is LHCf~\cite{lhcf:tdr}.  The primary objective of the LHCf collaboration is to measure the forward production spectra of photons and $\pi^{\circ}$'s in high-energy $pp$ collisions to constrain the models of high energy cosmic ray air shower development.  The LHCf detectors are tungsten/scintillator calorimeters with silicon microstrip and scintillating fibre trackers located at $\pm 140$~m from the interaction point.

The CMS~\cite{cms:tdr} experiment is the general purpose detector at IP5.  The TOTEM~\cite{totem:tdr} experiment shares IP5 with CMS.  The CMS forward detectors include HF, CASTOR, and CMS-ZDC, while the TOTEM detectors, T1/T2 and RP, are all in the forward region.  While TOTEM and CMS are independent collaborations, the experiments will have integrated read-out, allowing TOTEM to benefit from the CMS central coverage, and CMS to benefit from the TOTEM forward coverage.  

The CMS HF is a steel/quartz calorimeter located 11~m from the CMS IP ($3<|\eta|<5$).  The primary motivation for the HF is forward jet tagging for the vector-boson-fusion Higgs production channel, but it will play an important role in low-$x$ and forward physics as well.  CASTOR is a tungsten/quartz calorimeter with electromagnetic ($5.3<|\eta|<6.5$) and hadronic  ($5.2<|\eta|<6.4$) components located at $\pm14$~m from the CMS IP.  The primary objective of the CASTOR calorimeter in $pp$ collisions is the study of the proton PDFs at very low $x$ ($\sim 10^{-6}$) using Drell-Yan measurements.  The CMS-ZDC is a tungsten/quartz calorimeter at $\pm140$~m from the CMS IP.  Like the ATLAS-ZDC, the CMS-ZDC's primary physics objective is to measure the centrality in heavy-ion collisions.  In the $pp$ program it can be used to study charge exchange events with a leading neutron as well as forward energy flow.

The primary objectives of the TOTEM experiment are to measure; the $pp$ elastic cross section as a function of  the square of the exchanged four-momentum ($t$), the $pp$ total cross section with a precision of approximately 1\%, and diffractive dissociation in $pp$ collisions at $\sqrt{s} = 14$~TeV.  The TOTEM experiment consists of 2 tracking telescopes T1 and T2, as well as Roman Pot (RP) stations. The T1 ($3.2 < |\eta| < 5$) and T2 ($5 < |\eta| < 6.6$) telescopes consist of cathode strip chambers and gas electron multiplier chambers, respectively.  The TOTEM RP stations containing silicon strip detectors will be placed at a distance of $\pm147$~m and $\pm220$~m from IP5.  The stations can measure protons with a momentum loss $\xi = \Delta p/p$ in the range $0.02 < \xi < 0.2$ for the nominal collision optics.  For other optics with larger $\beta^*$, and hence lower luminosity, much smaller values of $\xi$ can be reached. 

The FP420 (\emph{Forward Protons at 420m})~\cite{fp420:loi} research and development project is aiming to instrument both IP1 and IP5 with 3D silicon trackers and fast timing proton taggers to detect the leading proton from high mass exclusive diffractive $pp$ interactions.  

\subsection{Physics Prospects}
\label{sec:physics}

\begin{figure}[htbp]
   \centering
   \includegraphics[width=3.5in]{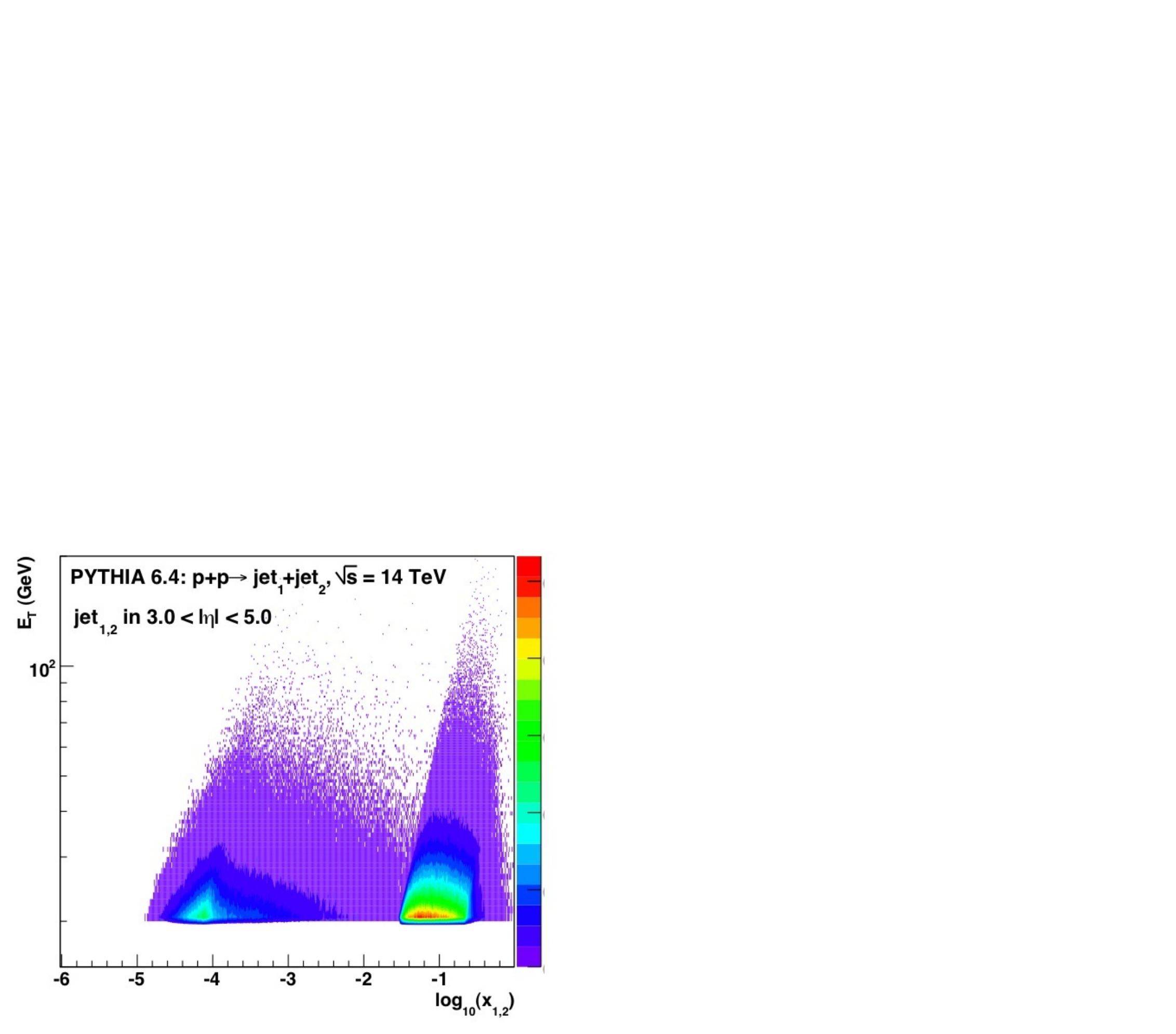} 
   \caption{Generator level  log($x_{1,2}$) distribution accessible at CMS given that one of the partons falls within the HF acceptance ($3<|\eta|<5$ and $E_T > 20$~GeV) }
   \label{fig:fwdjets}
\end{figure}

The high cross section of forward jet production makes it a favourable channel to study the low-$x$ behaviour of the proton PDF at 14~TeV.  Figure~\ref{fig:fwdjets} (based on PYTHIA 6.403~\cite{pythia64}) shows that $x$ as low as $10^{-4} - 10^{-5}$ will be accessible in the CMS HF.   So far, measurements at the Tevatron have probed the proton PDF down to $x\simeq 10^{-3}$.  The CMS analysis of forward jets has two primary objectives; the single inclusive jet cross-section in HF with $E_T \simeq 20 - 100$~GeV, and the "Muller-Navelet" (MN) dijet cross-section with a jet in each of the $+\eta$ and $-\eta$ HF detectors.  

\begin{figure}[htbp]
   \centering
   \includegraphics[width=2.5in]{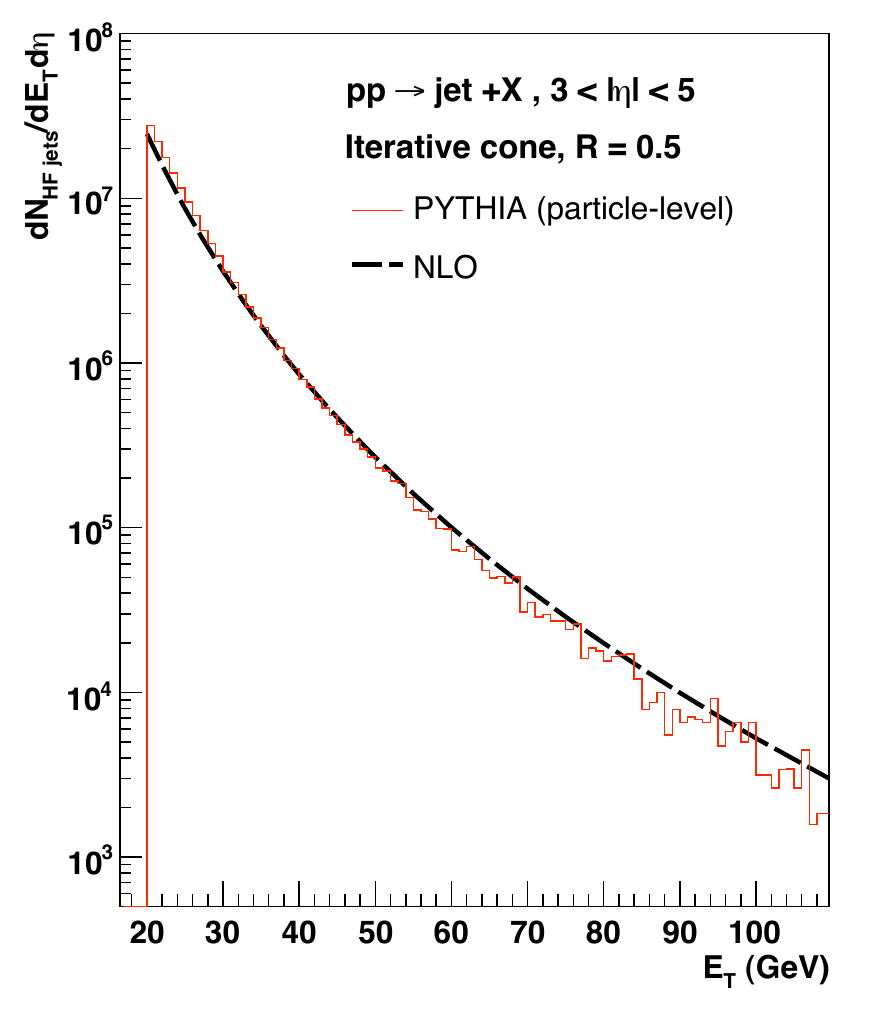} 
   \caption{Number of single inclusive jet events within the CMS HF acceptance expected in 1~pb$^{-1}$ of data  using particle-level jets (cone R=0.5) from PYTHIA 6.403 with no detector response, underlying event, or hadronization corrections applied.}
   \label{fig:jetcs}
\end{figure}

The primary objective of the single inclusive jet cross-section measurement is to constrain the low-$x$ proton PDF by including the data into the global PDF fits.  This initial analysis considers only 1~pb$^{-1}$ of low luminosity data, since pile-up (a potentially significant contribution to the systematic error) is negligible at low luminosity.  Figure~\ref{fig:jetcs} shows that the statistics available in just 1~pb$^{-1}$ are very large, but a full calculation of the systematic uncertainties (primarily from detector response, underlying event, hadronization, and luminosity) needs to be completed before a definitive statement on the possible constrains on the PDF are possible.

\begin{figure}[htbp]
   \centering
   \includegraphics[width=3.5in]{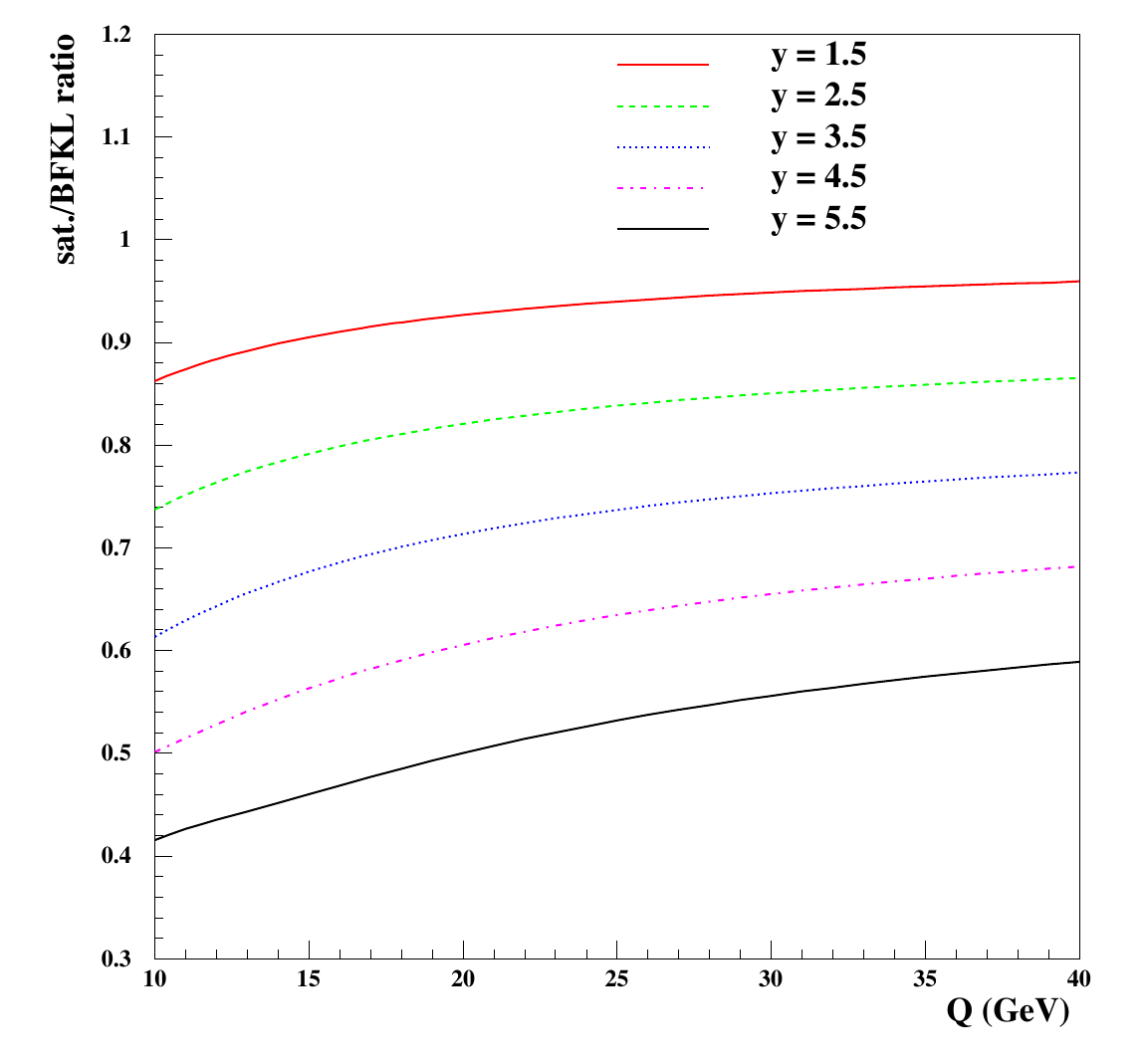} 
   \caption{Suppression of the MN forward dijet cross-section with low-$x$ saturation compared to BFKL as a function of $Q \equiv Q_1 = Q_2$.}
   \label{fig:mn}
\end{figure}

The MN dijet measurement is particularly sensitive to the BFKL~\cite{mueller:87} and small-$x$~\cite{marquet:06} evolution of the proton.  The colliding partons in the MN kinematics are both large-$x$ valence quarks ($x_{1,2}\approx0.1$) which produce a large rapidity interval between the two jets.  The large rapidity separation enhances the available phase space in longitudinal momentum for BFKL radiation.  Recent works~\cite{marquet:06} indicate that the presence of low-$x$ saturation effects will suppress the the forward-backward MN dijet production cross section compared to the BFKL prediction, as seen in Figure~\ref{fig:mn}.  By using the $+\eta$ and $-\eta$ HF calorimeters, $\Delta \eta \sim 9$ (where the suppression is $\sim$2 at low $Q$) can be achieved in the CMS detector.  A full study of the rate and jet reconstruction efficiency in the HF must be performed before a conclusion can be drawn regarding the feasibility of this measurement.

\begin{figure}[htbp]
   \centering
   \includegraphics[width=3.5in]{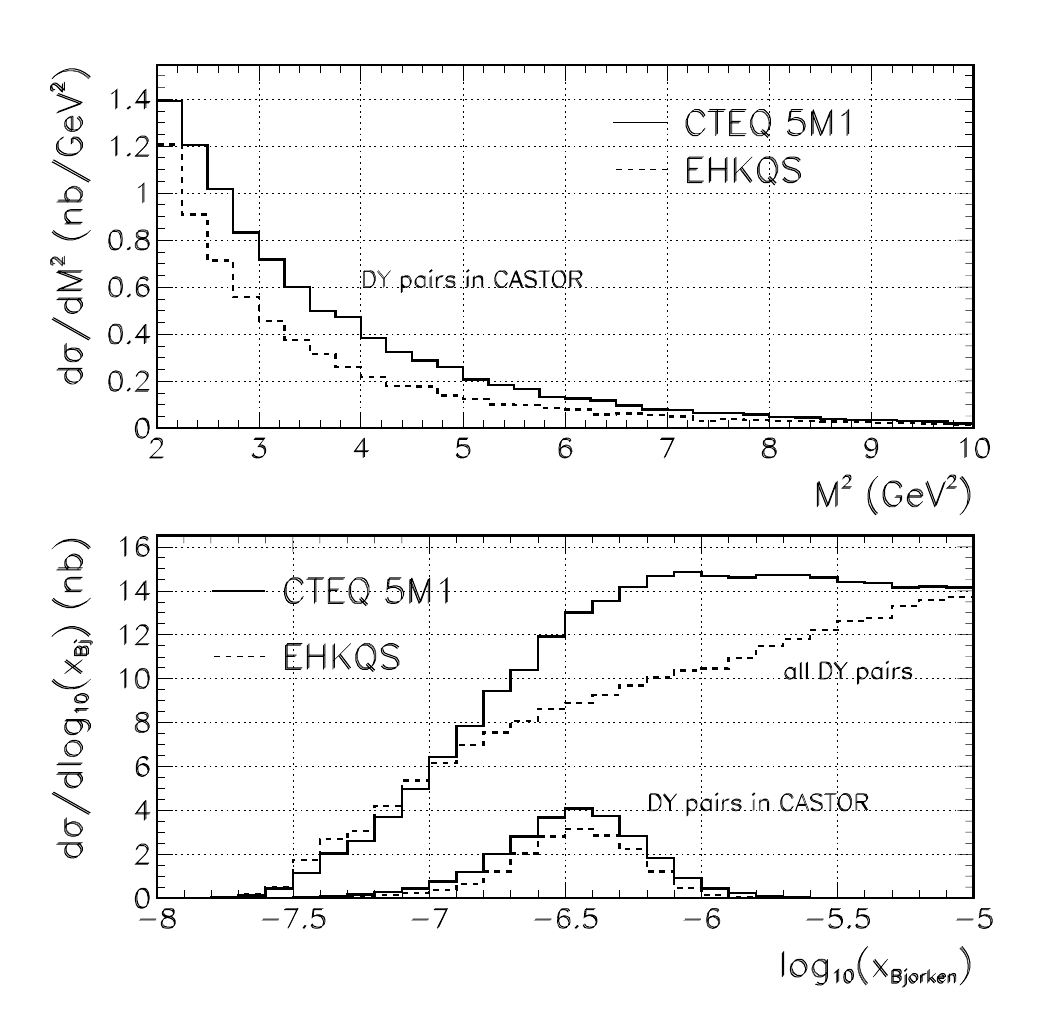} 
   \caption{Comparison Drell-Yan pairs of non-saturation (CTEQ) and saturation (EHKQS) models using the CMS CASTOR detector.}
   \label{fig:dy}
\end{figure}

Another physics process that could provide information on saturation effects are forward Drell-Yan pairs.  With a large imbalance between the  $x$ of the $q$ and $\bar{q}$, the Drell-Yan pair is boosted to large rapidities, and the low-$x$ $q$ and $\bar{q}$ distributions become accessible.  The  effect of saturation on the cross section is shown in Figure~\ref{fig:dy}, where a standard parameterisation of the parton density function (CTEQ 5M1) is compared to a ÒsaturatedÓ parameterisation (EHKQS)~\cite{ehkqs}. In the kinematic range accessible by the CMS CASTOR calorimeter, a 30\% decrease in the cross section is observed.  Again, more study is required to make a definitive statement on the feasibility of this measurement, but initial studies are very promising.

\begin{figure}[htbp]
   \centering
   \includegraphics[width=3.5in]{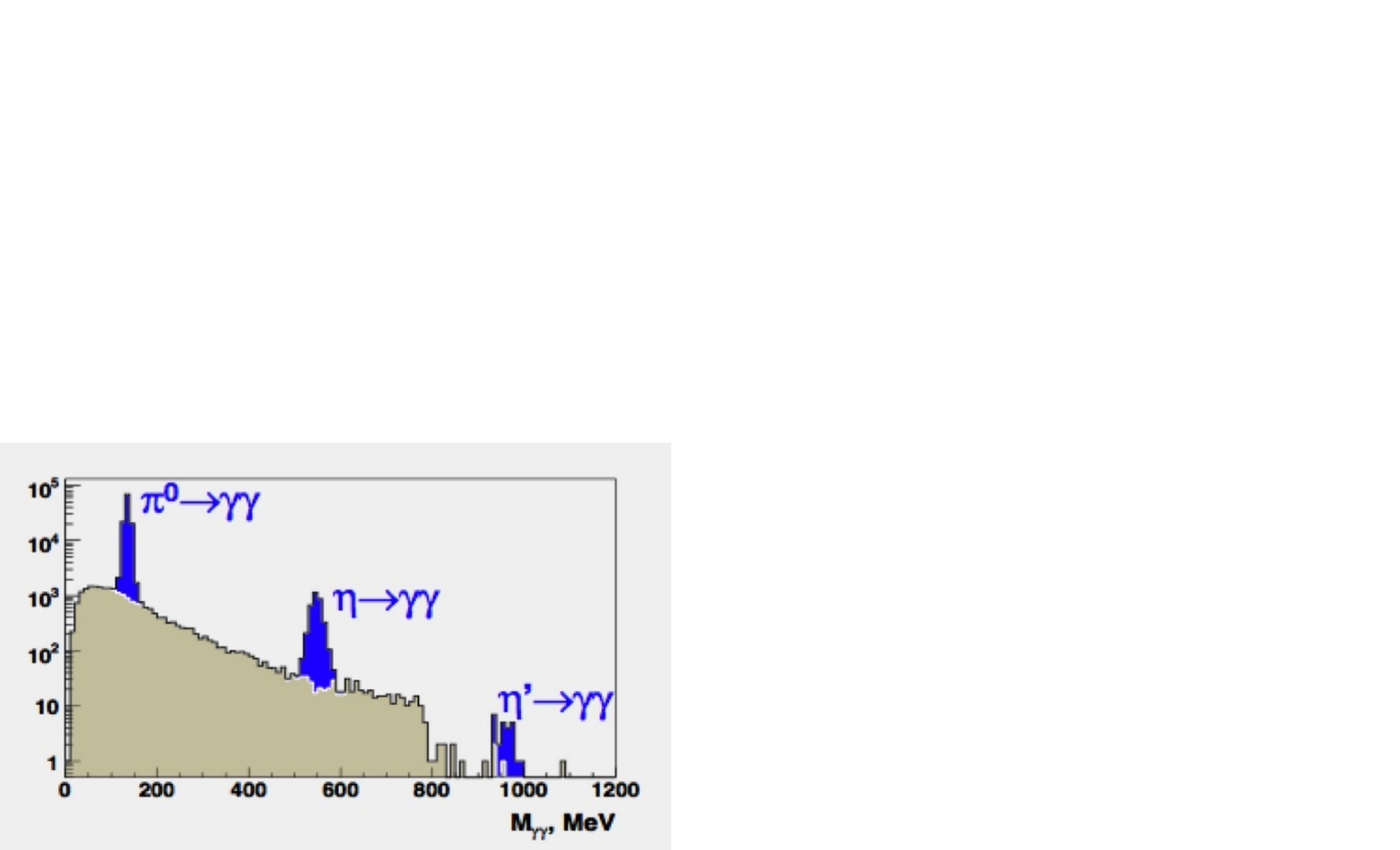} 
   \caption{Potential for forward neutral scalar boson reconstruction in the ATLAS ZDC}
   \label{fig:fwdparts}
\end{figure}

While the ATLAS and CMS ZDC's primary focus is in the heavy ion program, they have significant potential for measuring forward particle production in the $pp$ program as well.   The ability to tag a forward neutron allows for the possibility to look for processes like $p + \gamma \rightarrow W + n$, where the forward neutron is used to identify the charge exchange process.  The potential for this process to be used to study trilinear gauge boson couplings has not yet been studied with detector acceptances and efficiencies, but theoretical work is underway~\cite{dreyer:ect}.  Measuring very forward neutral scalar particle production  is also possible with the ATLAS and CMS ZDC's.  The reconstruction of $\pi^{\circ} \to \gamma \gamma$, $\eta \to \gamma\gamma$, and $\eta' \to \gamma\gamma$ is shown using the ATLAS ZDC in Figure~\ref{fig:fwdparts}.

\begin{figure}[htbp]
   \centering
   \includegraphics[width=6.5in]{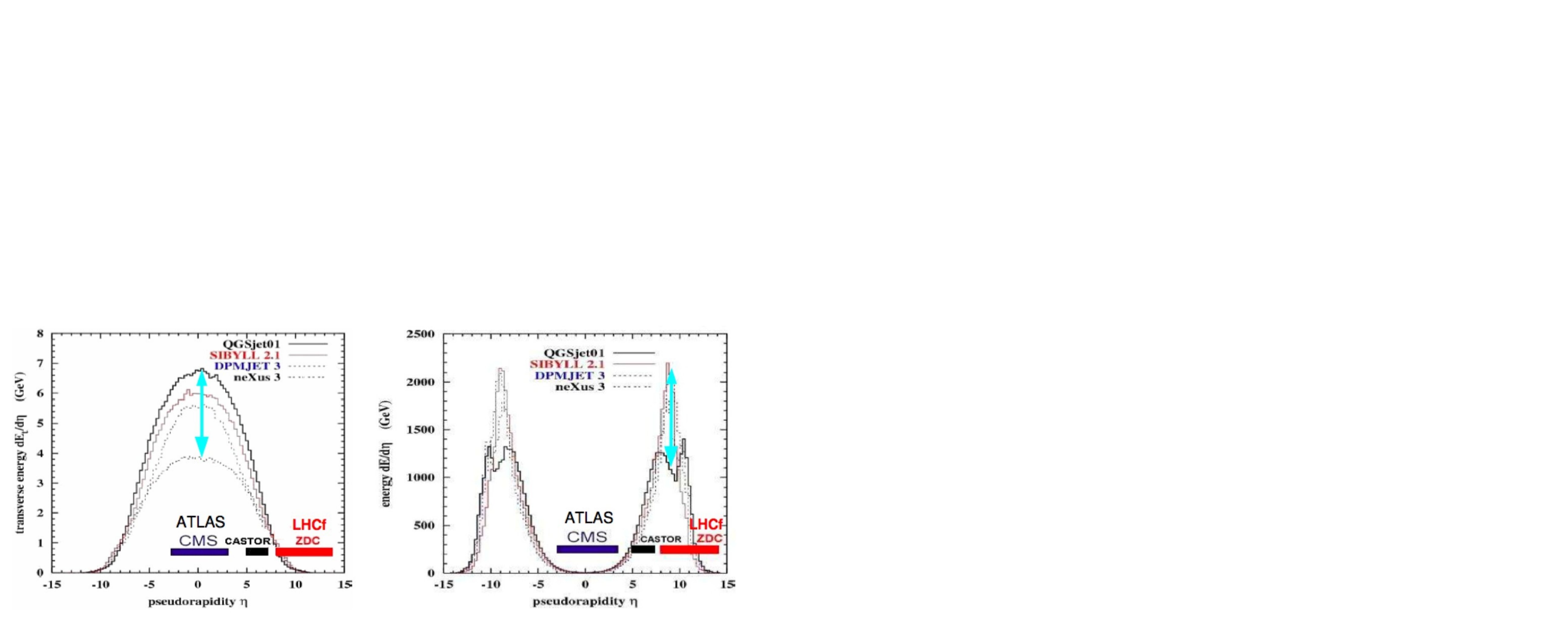} 
   \caption{Comparison of various high energy cosmic ray air shower simulations with an overlay of the acceptance of the ATLAS, CMS, and LHCf experiments}
   \label{fig:cosmic}
\end{figure}

The LHCf experiment will also reconstruct very forward $\pi^{\circ} \to \gamma\gamma$ events and measure their energy spectrum as well as providing information about forward energy flow for constraining high energy cosmic ray air shower simulations.  The modeling of high energy cosmic ray air showers is important to help resolve the apparent conflict in the results of the HiRes~\cite{hires:gzk} and AGASA~\cite{agasa:gzk} experiments regarding the GZK cutoff.  The predictions of various models are shown in Figure~\ref{fig:cosmic}.  The LHCf experiment will measure the forward production spectra of $\pi^{\circ}$'s and photon's to constrain these models to improve the shower simulation precision and help resolve the GZK cutoff question.

\subsection*{Conclusions}

The ATLAS, CMS, TOTEM, and LHCf experiments are well prepared to explore the low-$x$ and forward energy flow physics that will become accessible in the $\sqrt{s}$=14~TeV $pp$ collisions of the LHC.  Using a variety of detector systems and event signatures, the experiments will cover a broad spectrum of exciting forward physics measurements.

\newpage

\subsection*{Acknowledgments}

This talk is on behalf of the ATLAS, CMS, TOTEM, and LHCf collaborations.  I would firstly like to thank the members of all four collaborations for the huge effort putting these complex detectors together and developing their physics programs.  Specifically, I would like to thank David d'Enterria, Pierre Van Mechelen, Oscar Adriani, James Pinfold, Stephan Ask, Monika Grothe ,and Per Grafstrom from their contributions to these proceedings.




\begin{footnotesize}
\bibliographystyle{blois07} 
{\raggedright
\bibliography{LHCFwdPhys_blois07}
}
\end{footnotesize}
\end{document}